# AN OPTICAL AND TERAHERTZ INSTRUMENTATION SYSTEM AT THE FAST LINAC AT FERMILAB*

R. Thurman-Keup[†], A. H. Lumpkin, J. Thangaraj, FNAL, Batavia, IL, 60510, USA

*Abstract*

FAST is a facility at Fermilab that consists of a photoinjector, two superconducting capture cavities, one superconducting ILC-style cryomodule, and a small ring for studying non-linear, integrable beam optics called IOTA. This paper discusses the layout for the optical transport system that provides optical radiation to an externally located streak camera for bunch length measurements, and THz radiation to a Martin-Puplett interferometer, also for bunch length measurements. It accepts radiation from two synchrotron radiation ports in a chicane bunch compressor and a diffraction/transition radiation screen downstream of the compressor. It also has the potential to access signal from a transition radiation screen or YAG screen after the spectrometer magnet for measurements of energy-time correlations. Initial results from both the streak camera and Martin-Puplett will be presented.

## INTRODUCTION

The Fermilab Accelerator Science and Technology (FAST) facility has been constructed for advanced accelerator research [1-3]. It will eventually consist of 3 entities: a photoinjector-based linac followed by an ILC-type cryomodule, an RFQ-based proton injector, and a small ring called IOTA (Integrable Optics Test Accelerator) for studying non-linear optics among other things. Presently, the facility has the linac, cryomodule, and beamline to a high-energy dump. The IOTA ring is under construction and should be completed next year. The proton source is also under construction and will be completed in the next couple of years. When the full beamline to IOTA has been completed, the electron linac will provide beam to IOTA to map out the optics of the ring in preparation for injecting protons once the source is completed. In addition to the experiments in the IOTA ring, there are various experiments in the linac; both before the cryomodule and after. In support of these experiments, there is an optical / THz instrumentation system containing a a Hamamatsu streak camera and a Martin-Puplett interferometer. This paper will describe the system and present some initial measurements from it.

## FAST FACILITY

The FAST injector starts with a 1.3 GHz normal-conducting rf photocathode gun with a $Cs_2Te$ coated cathode. The photoelectrons are generated by irradiation with a YLF laser at 263 nm that can provide several μJ per pulse [4]. Following the gun are two superconducting 1.3 GHz capture cavities that accelerate the beam to its design energy of around 50 MeV. After acceleration, there is a section for doing round-to-flat beam transforms, followed by a magnetic bunch compressor and a short section that can accommodate small beam experiments. At the end of the experimental section is a spectrometer dipole which can direct the beam to the low energy dump. If the beam is not sent to the dump, it enters the ILC-type cryomodule where it receives up to 250 MeV of additional energy and is sent to either a high-energy dump or the IOTA ring. Table 1 lists the typical beam parameters.

Table 1: Beam Parameters for FAST

| Parameter | Value |
|---|---|
| Energy | 20 – 300 MeV |
| Bunch Charge | < 10 fC – 3.2 nC |
| Bunch Frequency | 0.5 – 9 MHz |
| Macropulse Duration | ≤ 1 ms |
| Macropulse Frequency | 1 – 5 Hz |
| Transverse Emittance | > 1 μm |
| Bunch Length | 0.9 – 70 ps |

## OPTICAL AND TERAHERTZ INSTRUMENTATION SYSTEM

The optical / THz instrumentation system (OTIS) [5] is located near the low energy dump. The transport system is constructed of stainless steel pipes and flanges connected between boxes containing mirrors for steering the light (Fig. 1).

There are currently three source points for the radiation. One each from the third and fourth dipoles in the bunch compressor, D3 and D4, and one from a cross downstream of the bunch compressor, X121. There is also the possibility of implementing one from the cross (X124) in the low energy dump beamline that normally measures the energy spread.

The source points D3 and D4 in the bunch compressor provide coherent synchrotron edge radiation (CSR) from the entrance of the corresponding dipole magnets. The CSR is generally in the high GHz / low THz region [6].

The source point at X121 consists of an aluminized, silicon wafer on a translatable actuator (Fig. 2). The wafer can be inserted all the way into the beam to generate both coherent transition radiation (CTR) and optical transition radiation (OTR). The screen can also be extracted partly, with its edge a few mm above the beam, to generate coherent diffraction radiation (CDR).

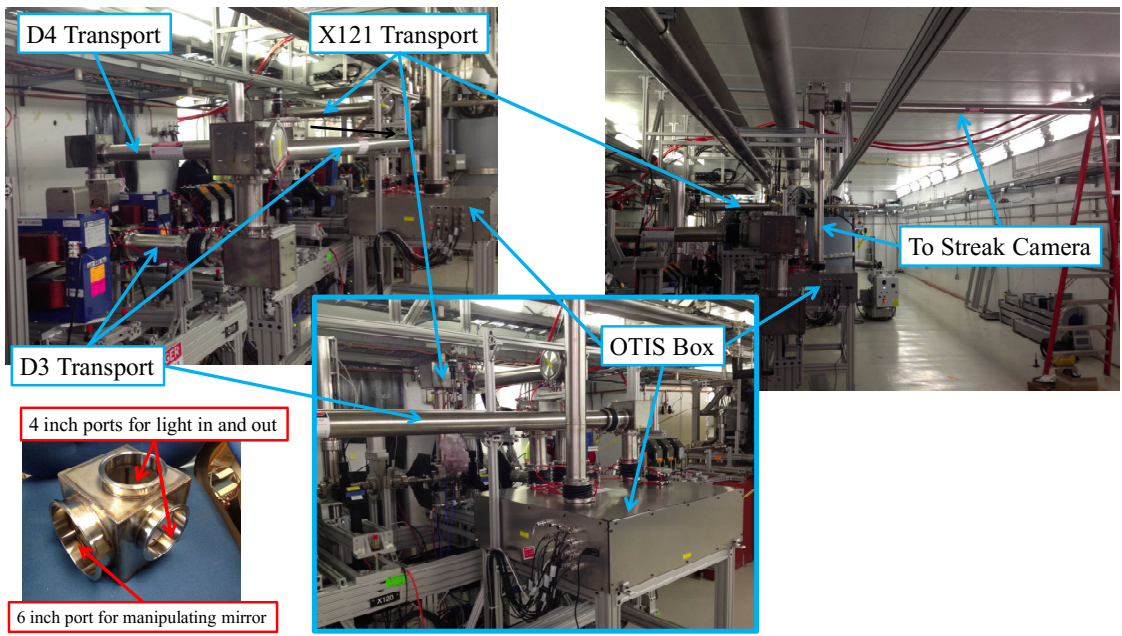

Figure 1: OTIS transport system. The transport pipes run between boxes (lower left) containing 4 inch flat mirrors.

All the sources converge to the OTIS box containing translatable mirrors to switch between the various inputs (Fig. 3). There are effectively three outputs from the box: a Hamamatsu streak camera, a Martin-Puplett interferometer, which is housed inside the OTIS box, and a 'user' location in the box which is currently unpopulated. Any input can be switched to any output. In addition, the mirror to switch light to the streak camera is a dielectric bandpass mirror with a reflection bandwidth of 400 – 700 nm. This transmits GHz/THz radiation to the Martin-Puplett interferometer.

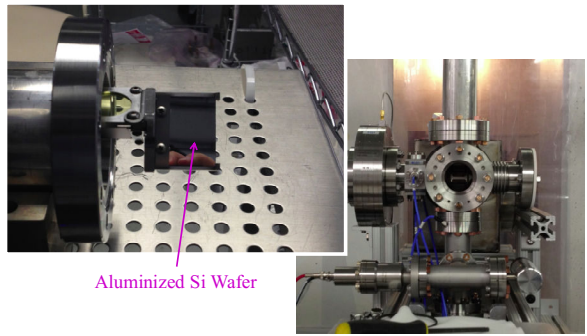

Figure 2: Left) X121 actuator retracted; OTR wafer is visible. Right) X121 cross with actuator installed.

The transport lines consist of stainless steel, 4 inch diameter pipes with quick disconnect flanges (Tri-clamp). In the bunch compressor, light is extracted through a single crystal quartz viewport. Since space is limited at these ports, a small box with one or two flat mirrors is attached to the port to redirect the radiation to a more readily accessible location. The radiation is then collimated by a pair of 3 inch diameter, 90° off-axis, parabolic, aluminium coated, UV-enhanced mirrors located in stainless steel boxes. A pair is needed since the focal length of the mirrors is much smaller than the closest distance to the dipole source point. The mirrors are mounted in an adjustable mirror mount which sits on a translatable stage to enable both direction and focusing adjustments. For X121, the light is also extracted through a single-crystal-quartz viewport. Here it is not necessary to have two focusing mirrors, since the mirror can be positioned at its focal point.

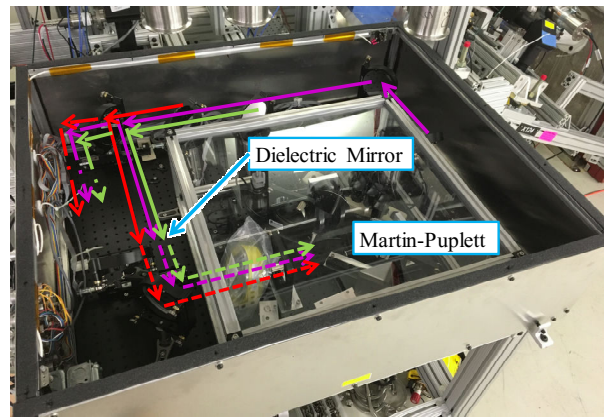

Figure 3: Light path inside the OTIS box. The colors indicate source, Purple – D3 in bunch compressor, Red – D4 in bunch compressor, Green – X121. Solid lines go to the streak camera. Dashed lines go to the interferometer, Dot-Dash lines go to the user area in the box.

Ninety-degree bends in the transport are handled by 4 inch diameter, flat, aluminium-coated, UV-enhanced mirrors mounted in an adjustable mount. To enable the system to be disassembled when access to the beamline is needed, the longer pipes are cut in two and have a rubber bellows between them.

The streak camera is located outside the accelerator enclosure. The pipe running to the streak camera goes through one of the penetrations to the outside where the streak camera is housed in a small enclosure.

## ALIGNMENT

The OTIS setup has many mirrors which forces one to consider how to align everything such that light gets from one end to the other. For the D3 and D4 ports, a laser is directed down the beamline with its waist at the location of the radiation source point. The mirrors are adjusted first for directionality and then for focus with the process iterated as necessary. At X121, a permanent alignment laser upstream is used to set the OTR screen angle, while a local laser is used to set the transport alignment and focusing. Figure 4 shows the local alignment laser.

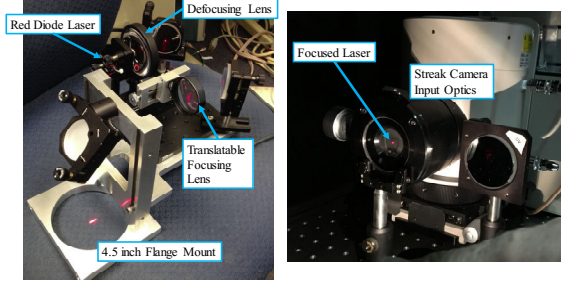

Figure 4: Left) Local alignment laser. It is designed to attach to a 4.5 inch Conflat flange. The focusing lens is on a linear stage to adjust the laser waist and provide a proper source point for focusing adjustments in the transport line. Right) Focused laser spot on the input slit of the streak camera after ~14 meters and 12 mirrors.

## INSTRUMENTATION

The end points of the optical and THz radiation include a Hamamatsu streak camera and a Martin-Puplett interferometer, both of which are used for bunch length measurements.

### Streak Camera

The streak camera [7] is a Hamamatsu C5680 mainframe with S20 PC streak tube that can accommodate a vertical sweep plugin unit and a horizontal sweep unit or blanking unit. The device is fitted with all-mirror input optics enabling the assessment of the UV OTR component as well as the visible light OTR. The mirror optics also mitigate the streak image blurring due to the inherent chromatic temporal dispersive effects of the lens-based input optics for broadband sources such as OTR. The unit is equipped with the M5675 synchroscan unit with its resonant circuit tuned at 81.25 MHz such that the streak image has jitter of less than 1 ps from the system itself.

### Interferometer

The Martin-Puplett interferometer is a polarizing-type interferometer with wire grids as the polarizer and beam splitters (Fig. 5). The wire grids are 10 μm diameter Tungsten wires with a 45-μm wire spacing. The device uses pyroelectric detectors to measure GHz/THz frequencies needed to determine the bunch length. The pyroelectric detectors are mounted on remote-controlled 2-D linear stages to allow optimization of the detected signal.

## TRANSPORT SIMULATION

The transport of GHz and THz frequencies over long distances is problematic due to the combined effect of diffraction and apertures. Simulations have been done for CTR/CDR at a frequency of 250 GHz. The simulation is written in MATLAB. It calculates the near-zone transition radiation from the X121 source using the prescription of [8]. The electric fields are evaluated at the first mirror location in the transport line and propagated from there using the vector diffraction formulation in [9]. The propagation presently takes into account the aperture of the mirrors, but not the change in direction. The intensity distributions at various points along the transport are shown in Fig. 6.

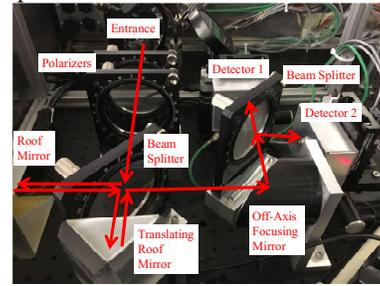

Figure 5: Interferometer light path. The GHz/THz is split and recombined from different path lengths. The interference determines the frequency content which in turn determines the bunch length.

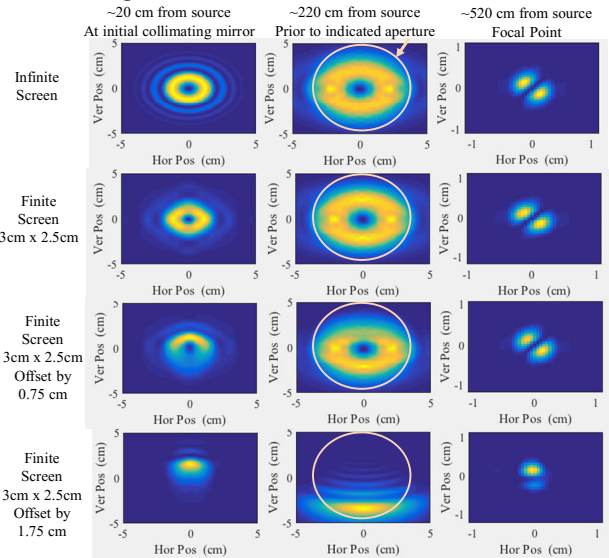

Figure 6: Simulated images along the transport line for several CTR screen scenarios. The bottom row is with the CTR screen withdrawn past the beam and functioning as a CDR screen. The asymmetric source point in this last scenario causes the light to propagate off-axis and impinge more severely on the aperture restrictions.

The apertures of the mirrors cause a loss of intensity down the transport line. The relative loss varies with CTR screen position. This effect can be understood qualitatively from the images where the peak intensities shift from the centered CTR screen version. This shift causes more light to be lost to the apertures. Figure 7 quantifies the loss down the transport line for various CTR screen positions.

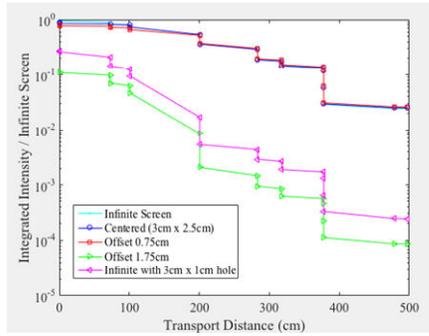

Figure 7: Integrated intensity along the transport line for a variety of CTR screen scenarios. The vertical steps in the intensity are due to either mirror apertures or polarizers. The offset of 1.75 cm corresponds to the screen acting as a CDR source.

## INITIAL DATA

The system has been used to measure the bunch length as a function of rf chirp induced by the second accelerating cavity (CC2). Figure 8 shows the rms bunch length as measured by the streak camera as a function of degrees off crest (DOC) of the rf accelerating gradient in CC2. The minimum bunch length is a bit under 1 ps. Other streak camera measurements can be found in [10].

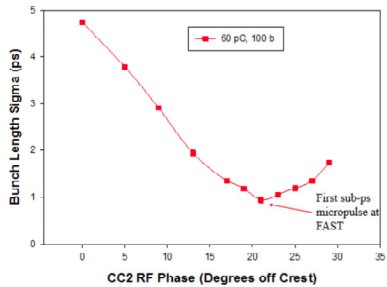

Figure 8: Streak camera measurements of rms bunch length using OTR from the X121 source point [11].

Initial setup of the Martin-Puplett involved scanning the pyroelectric detectors to find the peak signal. Figure 9 shows the measured and simulated signals. The interferometer has been used to make several autocorrelation measurements at varying bunch charges (Fig. 10). The bunch lengths were too long to distinguish subtle differences in length for different bunch charges, however the presence of an autocorrelation trace with a reasonable spectrum is promising for future measurements.

## SUMMARY

A general purpose optical and terahertz instrumentation system has been installed and partially commissioned at the FAST facility at Fermilab. Measurements have been taken utilizing the streak camera, and some initial measurements have been done using the Martin-Puplett interferometer. Further commissioning and use is expected in experiments planned for the next several months.

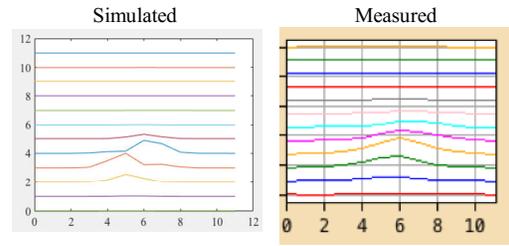

Figure 9: Simulated and measured scans of pyroelectric intensity. The pyroelectric detector is 2 mm x 2 mm. These scans are approximately 20 mm x 20 mm so each line is composed of ~10 4-mm$^2$ pixels. There is reasonably good agreement between measurement and simulation.

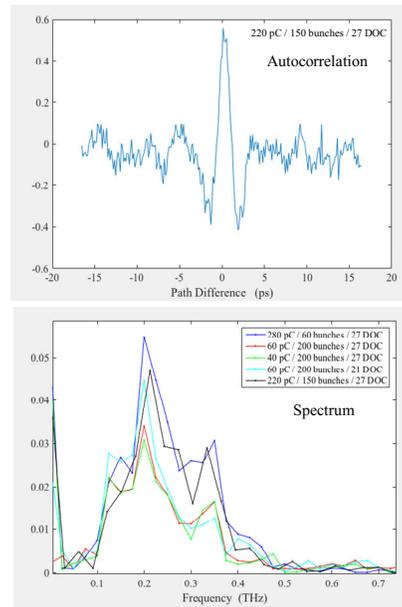

Figure 10: Top) Autocorrelation trace of CTR from electron beam near maximum compression. Bottom) Spectra reconstructed from autocorrelation data for myriad bunch charges and # of bunches. The spectra are approximately what one would expect for an rms bunch length near 1 ps.

## ACKNOWLEDGEMENTS

The authors would like to thank the FAST department staff for their efforts in constructing and installing this system. This manuscript has been authored by Fermi Research Alliance, LLC under Contract No. DE-AC02-07CH11359 with the U.S. Department of Energy, Office of Science, Office of High Energy Physics. The United States Government retains and the publisher, by accepting the article for publication, acknowledges that the United States Government retains a non-exclusive, paid-up, irrevocable, world-wide license to publish or reproduce the published form of this manuscript, or allow others to do so, for United States Government purposes.

## REFERENCES

[1] P. Garbincius *et al*., "Proposal for an accelerator R&D user facility at Fermilab's Advanced Superconducting Test Ac-

celerator (ASTA)", FNAL, Batavia, IL, USA, Rep. Fermilab-TM-2568, October 2013.

[2] S. Antipov *et al*., "IOTA (Integrable Optics Test Accelerator): facility and experimental beam physics", *Journal of Instrumentation*, vol. 12, p. T03002, 2017.

[3] D. Edstrom, Jr. *et al*., "50-MeV run of the IOTA/FAST electron accelerator", in *Proc. NAPAC'16*, Chicago, IL, USA, October 2016, paper TUPOA19.

[4] J. Ruan *et al*., "Commission of the drive laser system for advanced superconducting test accelerator", in *Proc. IPAC'13*, Shanghai, China, May 2013, paper WEPME057.

[5] R. Thurman-Keup *et al*., "Terahertz and optical measurement apparatus for the Fermilab ASTA injector", in *Proc. IBIC'14*, Monterrey, CA, USA, September 2014, paper TUPD03.

[6] J. C. T. Thangaraj *et al*., "Experimental studies on coherent synchrotron radiation at an emittance exchange beam line", *Phys. Rev. ST Accel. Beams* vol. 15, p. 110702, 2012.

[7] A. H. Lumpkin *et al*., "Commissioning of a dual-sweep streak camera with applications to the ASTA photoinjector drive laser", in *Proc. FEL'14*, Basel, Switzerland, August 2014, paper MOP021.

[8] A. G. Shkvarunets and R. B. Fiorito, "Vector electromagnetic theory of transition and diffraction radiation with application to the measurement of longitudinal bunch size", *Phys. Rev. ST Accel. Beams* vol. 11, p. 012801, 2008.

[9] T. J. Maxwell *et al*., "Vector diffraction theory and coherent transition radiation interferometry in electron linacs", in *Proc. PAC'07*, Albuquerque, NM, USA, paper FRPMS035.

[10] A. H. Lumpkin *et al*., "Initial demonstration of 9-MHz framing camera rates on the FAST drive laser pulse trains", in *Proc. NAPAC'16*, Chicago, IL, USA, October 2016, paper TUPOA25.

[11] A. H. Lumpkin *et al*., "Initial observations of micropulse elongation of electron beams in a SCRF accelerator", in *Proc. NAPAC'16*, Chicago, IL, USA, October 2016, paper TUPOA26.